\documentclass[final,5p,times,twocolumn]{elsarticle}
\usepackage{amsmath}
\journal{Computer Physics Communications}

\begin{document}


\begin{frontmatter}

\title{The DEPOSIT computer code: calculations of
electron-loss cross sections  for\\
complex ions colliding with neutral atoms.}

\author[uuse]{Mikhail S. Litsarev}
\ead{mikhail.litcarev@physics.uu.se}
\address[uuse]{Department of Physics and Astronomy, Uppsala University,
Box 516, 75120 Uppsala, Sweden}

\begin{abstract}
A description of the DEPOSIT computer code is presented. The code
is intended to calculate \textit{total} and $m$\textit{-fold}
electron-loss cross sections ($m$ is the number of ionized
electrons) and the energy $T(b)$ deposited to the projectile
(positive or negative ion) during a collision with a neutral atom
at low and intermediate collision energies as a function of the
impact parameter $b$. The deposited energy is calculated as a
3D-integral over the projectile coordinate space in the classical
energy-deposition model. Examples of the calculated deposited
energies, ionization probabilities and electron-loss cross
sections are given as well as the description of the input and
output data.
\end{abstract}


\end{frontmatter}

\today

\section{Program summary}
\begin{description}
\item[Program title:] DEPOSIT
\item[Licensing provisions:] GNU General Public License version 3
\item[No. of lines in distributed program, including test data, etc.:]
\item[Distribution format:] tar.gz 
\item[Programming language:]C++
\item[Computer:] Any computer that can run C++ compiler.
\item[Operating system:] Any operating system that can run C++ compiler. 
\item[Classification:] 2.4, 2.6, 4.10, 4.11
\item[Nature of problem:] For a given impact parameter $b$ to calculate the
deposited energy $T(b)$ as a 3D integral over a coordinate space,
and ionization probabilities $P_m(b)$.

For a given energy to calculate the total and $m$-fold
electron-loss cross sections using $T(b)$ values.

\item[Solution method:] Direct calculation of the 3D-integral $T(b)$.
One-dimension quadrature formula of the highest accuracy based upon
the nodes of the Yacobi polynomials for the $\cos \theta=x \in
[-1,1]$ angular variable is applied. The Simpson rule for the $\varphi \in
[0,2\pi]$ angular variable is used. Newton-Cotes pattern of the
seventh order embedded into every segment of the logarithmic
grid for the radial variable $r \in [0,\infty]$ is applied.
Clamped cubic spline interpolation is done for the 
integrand of the $T(b)$.

Bisection method and further parabolic interpolation
is applied for the solving of the nonlinear equation
for the total cross-section.
The Simpson rule for the $m$-fold cross-section
calculation is applied.

\item[Running time:] For a given energy, the total and $m$-fold
cross sections are calculated within about $15$ minutes on
$8$-core system. The running time is directly proportional to the
number of cores.
\end{description}

\section{Introduction.}
\label{IntroSect} Electron loss (also called projectile ionization
or stripping) of positive and negative ions, colliding with
neutral atoms, is an important collision process playing a key
role in many problems of plasma physics
\cite{Hofmann1}-\cite{HofmanI2} and physics of accelerators
\cite{GSI6rep}-\cite{Mueller2}. For example, electron-loss (EL)
cross sections are required for estimation of the ion-beam
lifetimes and vacuum conditions (residual-gas pressure and
concentrations) in accelerators and storage rings, for energy
losses of high-temperature plasmas and for heavy-ion beam-probe
(HIBP) diagnostics of plasmas in tokamaks and stellarators.
General properties of various atomic processes occurring in the
collisions of heavy ions with neutral atomic and molecular targets
are investigated in detail and presented in various review
articles and books \cite{Betz}-\cite{Voitkiv}.

At low and intermediate energies, a multiple-electron losses
contribution to the EL cross sections can be very
large~\cite{Olson1}-\cite{Pirumal} (up to~50\% and more), but with
the energy increasing a contribution of multiple EL processes goes
down and single-electron losses become dominate. At high and
relativistic energies the EL cross sections can be calculated in
the relativistic Born
approximation~\cite{Shevelko2}-\cite{Shevelko3}. At low and
intermediate energies, the total EL cross sections can be
calculated in the classical approximation using the CTMC
approach~\cite{Olson2}, \cite{OlsonCTMC}-\cite{OlsonBook}
(Classical Trajectory Monte Carlo) or the energy-deposition model
\cite{Paper1}-\cite{Paper3}.

In the CTMC approach all projectile and target electrons are
considered as classical particles described by a system of a few
hundred coupled Newton-type first-order differential equations
with a few thousand trajectories. The corresponding computer code
is quite complicated in use and takes a very long run-time.

The DEPOSIT program was created as an alternative code to the CTMC
one and is intended to calculate the energy $T(b)$ deposited
as a function of impact parameter $b$
to the projectile (positive or negative \textit{ion}) during a
collision with a neutral atom. The obtained $T(b)$ values are then
used for total and $m$-fold EL cross-section
calculations (see Section~\ref{PhysModel1}). The code is based on the classical
energy-deposition model suggested by N. Bohr \cite{bohr} and
developed later in \cite{Cocke1}-\cite{Kabachnik} for multiple-electron ionization of \textit{atoms} by positive ions. In the DEPOSIT program, the
problem of obtaining the energy deposited to the \textit{ions} by
neutral atoms is reduced to the calculation of a 3D-integral over
the ion coordinate space. The code requires much less computer
time and provides approximately the same accuracy of the EL
cross sections as the CTMC method, i.e. a factor of 2.

In this paper we describe a numerical implementation of the
energy-deposition model in the DEPOSIT computer code and give a
brief user guide. One important property of the code is that it
allows one to simulate electron-loss cross sections for an
arbitrary projectile (neutral atom, positive or negative ion)
colliding with arbitrary atomic target. The validity conditions of
the code are also discussed.

The article is organized as follows. In Section \ref{PhysModel1}
we briefly describe basic ideas and physical points of the developed
method. In Section \ref{NumericalProc2} we use these results and
explain the numerical implementation of the model. Some basic
examples on how to use the DEPOSIT code from the user's point of
view are set out in Section \ref{ExamplesIO3}. Finally, conclusions are
given in Section \ref{ResultsDiscSect}. Atomic units are used
throughout the paper.

\section{Physical model}
\label{PhysModel1}

The energy $T(b)$ transferred to the projectile ion by a neutral
target atom as a function of the impact parameter $b$ is given by
a 3D-integral over the projectile coordinate space \cite{Paper3}:
\begin{equation}
\label{Tb3DIntegral}
T(b)=\sum_{\gamma} \int \rho_{\gamma}(r)
\,\Delta E_{\gamma}(p)\, d^{3} \mathbf{r}.
\end{equation}
In the sum index $\gamma$ labels principal~$n$ and orbital~$l$
quantum numbers of atomic shell, $\gamma=nl$. The summation is
over all occupied shells of the projectile ion.

In the DEPOSIT code, the projectile electron density
$\rho_{\gamma}(r)$ for the $\gamma$-shell is presented by the
nodeless Slater wave function
\begin{equation}
\label{rhodencity}
\rho_{\gamma}(r)=N_{\gamma}
P^{2}_{\gamma}(r), \qquad
P_{\gamma}(r)=\left[ \frac{(2\beta)^{2\mu+1}}
{\Gamma(2\mu+1)} \right]^{1/2}r^{\mu}e^{-\beta r},
\end{equation}
and normalized to the total number of the projectile electrons $N$
\begin{equation}
\sum_{\gamma}\int_{0}^{\infty} \rho_{\gamma}(r) \,dr =
\sum_{\gamma}N_{\gamma}= N.
\end{equation}
Here $\Gamma(x)$ is the Gamma function,
$\beta$ and $\mu$ are Slater
parameters~\cite{SlaterBook}, \cite{ShevSlaterPrm2001}.

In the classical approximation the kinetic energy, deposited to
the shell $\gamma$ by the target atom can be expressed in the
form:
\begin{equation}
\Delta E_{\gamma}=
\frac{(u_{\gamma} + \Delta u_{\gamma})^2}{2} - \frac{u^2_{\gamma}}{2},
\end{equation}
where $u_{\gamma}$ is the orbital velocity of the electron in the
$\gamma$-shell and $\Delta u_{\gamma}$ denotes the energy gain by
the electron. The values $u_{\gamma}$  and $\Delta u_{\gamma}$ are
obtained from the classical equations
\begin{equation}
u_{\gamma}=\sqrt{2I_{\gamma}},
\end{equation}
\begin{equation}
\label{Ederivative}
\Delta u_{\gamma}=
\int_{-\infty}^{\infty} \frac{dU(R)}{dp} dt=
\int_{-\infty}^{\infty} \frac{dU(R)}{dR}\frac{dR}{dp}dt,
\end{equation}
where $I_{\gamma}$ is the binding energy of the projectile
$\gamma$-shell. The function $U(R)$ denotes the field  of a
neutral atom at a distance~$R$ from its nucleus and is calculated
as a sum of three Yukawa-type potentials with five fitting
parameters $A_i$, $\alpha_i$ obtained from the
Dirac-Hartree-Fock-Slater calculations \cite{DHFS_fitt}:
\begin{equation}
\label{DiracUr}
U(R)=-\frac{Z}{R}\sum_{i=1}^{3}A_i e^{-\alpha_i R},
\qquad \sum_{i=1}^{3}A_i=1,
\end{equation}
\begin{equation}
R^2=p^2 + v^2_r t^2, \qquad
v_r=\sqrt{v^2 + u^2_{\gamma}}.
\end{equation}
Here $v$ denotes the velocity of the projectile
(the straight-line trajectory approximation is used)
and $v_r$ is the relative velocity.

The impact parameter $\mathbf{p}$ is a function of $\mathbf{b}$
and $\mathbf{r}$ and in spherical coordinates ($r$, $\theta$,
$\varphi$) has the following dependence in the frame of the moving
projectile:
\begin{equation}
\label{pbtetphiBound} p^2=(b-r\cos\theta)^2 + (r\cos\varphi \sin
\theta)^2.
\end{equation}

The energy gain $\Delta u_{\gamma}$ in eq. (\ref{Ederivative})
diverges at small impact parameters $p$ and must be 
renormalized~\cite{Paper3}.
After normalization of $\Delta u_{\gamma}$ the 
deposited energy $\Delta E_{\gamma}$ is finally
determined depending on the relation between ion velocity $v$ and
orbital velocity $u_{\gamma}$ as
\begin{equation}
\label{DeltEmainTheta}
\Delta E_{\gamma}(p)=
\Delta E^{<}_{\gamma}(p) \theta(u_{\gamma} - v)+
\Delta E^{>}_{\gamma}(p) \theta(v - u_{\gamma}),
\end{equation}
where
\begin{equation}
\label{EfinalLow}
\Delta E^{<}_{\gamma}(p)=
\frac{2N_{\mbox{\footnotesize{eff}}}^{(\gamma)}u_{\gamma}
\sum^{3}_{i=1}A_i F(\alpha_i p)}
{v_{r}\left(p+\frac{4 u_{\gamma}}{v_r v^2}  \right)},
\end{equation}
\begin{equation}
\label{EfinalHigh}
\Delta E^{>}_{\gamma}(p)=
\frac{2Z^2  \left[\sum_{i=1}^3 A_i F(\alpha_i p) \right]^2 }{v^2(p^2 + Z^2/v^4)}.
\end{equation}
The step theta-function $\theta(x)$ is defined in a standard way
\begin{equation}
\label{StepTheta}
\theta(x)=
\begin{cases}
0, & x < 0\\
1/2, & x = 0\\
1, & x > 0.
\end{cases}
\end{equation}
In order to avoid the thresholds in every point
$v=u_{\gamma}$ due to the discrete behaviour of the $\theta(x)$
(i.e. to make the deposited energy $\Delta E_{\gamma}(p)$ a smooth function),
the step theta-function is replaced by the right-hand side Fermi-smearing function
\begin{equation}
\label{kSmear}
n_{f}(x)=\frac{1}{e^{-k x}+1}.
\end{equation}
The $n_{f}(x)$ becomes the $\theta(x)$ when $k \to \infty$.
Parameter value $k = 3$ is taken by default.
It is an input parameter and can be changed, if it is needed.
Thus,
\begin{equation}
\label{DeltEmainTheta}
\Delta E_{\gamma}(p)=
\Delta E^{<}_{\gamma}(p) \,n_{f} (u_{\gamma} - v)+
\Delta E^{>}_{\gamma}(p) \,n_{f} (v - u_{\gamma}),
\end{equation}

The constant variable $N_{\mbox{\footnotesize{eff}}}^{(\gamma)}$
depends on the mean radius $r_{\gamma}$ of the projectile shell
and on the atomic radius $R_A$ of the target atom
\begin{equation}
\label{NeffCase}
N_{\mbox{\footnotesize{eff}}}^{(\gamma)}=
\begin{cases}
Z\frac{r^{2}_{\gamma}}{R^{2}_{A}}, \quad & r_{\gamma} \le R_{A}, \\
Z, \quad & r_{\gamma} > R_{A},
\end{cases}
\end{equation}
where $r_{\gamma} = \mu_{\gamma}I^{-1/2}_{\gamma}$. The mean
radii $R_A$ for neutral atoms are given in \cite{Desclaux}.
The validity conditions are discussed in \cite{nimb09}.

The function $F(y)$ in eqs. (\ref{EfinalLow})
and~(\ref{EfinalHigh}) is a result of using the potential $U(R)$
in the form (\ref{DiracUr}) and is given by
\begin{equation}
\label{FyDef}
F(y)=\int_{0}^{\infty}  y^2 e^{-\sqrt{y^2 + x^2}} \left[
\frac{1}{(y^2+x^2)^{3/2}}+ \frac{1}{y^2+x^2} \right]dx.
\end{equation}
It can also be represented as
\begin{equation}
\label{FinalFyExpr}
F(y)=y K_1(y)=
\int_{0}^{\infty}
e^{-\sqrt{y^2 + x^2}}  d x,
\end{equation}
where $K_1(y)$ is the \textit{modified Bessel function of the
second kind}~(see \cite{Abramowitz}, p.~375, eq.~9.6.2, \cite{Gradshteyn}).
Integral~(\ref{FinalFyExpr}) does converge for all positive $y$.
It has the following expansion for $y$ close to zero
(which can be obtained from ref.~\cite{Abramowitz},
eq.~9.8.7,  p.~379 and eq.~9.8.3, p.~378)
\begin{equation}
F(y \to 0_+)=
1+\left(\frac{1}{2}\ln\frac{y}{2}+0.03860786\right)y^2+O(y^3).
\end{equation}
To speed up the calculations 
it is interpolated by using of clumped cubic 
splines~\cite{StoerSplines}, p.~97.

Once the $T(b)$ energy is determined, the \textit{total} EL
cross-section can be obtained directly as
\begin{equation}
\label{sigmatotEq}
\sigma_{tot}(v)=\sum_{m=1}^{N} \sigma_m = \pi b_{\max}^{2},
\end{equation}
where $b_{\max}$ is (numerically) found from the equation
\begin{equation}
\label{bmaxEquation}
T(b_{\max})=I_{1}
\end{equation}
and $I_1$ denotes the (first) ionization potential of the
projectile.

The $m$-fold EL cross-section $\sigma_m(v)$ can be calculated as
\begin{equation}
\label{sigmam}
\sigma_m(v)=2\pi \int^{\infty}_{0} P_m(b) b\, db,
\end{equation}
where $P_m(b)$ is a probability to ionize $m$ electrons from the
projectile.

In the CTMC method the probabilities $P_m(b)$ are
calculated numerically. To calculate $P_m(b)$ in the present
approach the Russek-Meli model \cite{Russek2} is used.
In this model the $P_m(b)$ probabilities are given in a close
analytical form as functions of the deposited energy $T(b)$
and the ionization potentials~$I_{k}$ of the projectile
\begin{equation}
\label{PmbRussekMeli}
P_m(b)=\binom{N'}{m}\, S_{m}(\eta_m(b))
/\sum_{i=1}^{N'} \binom{N'}{i} \, S_{i}(\eta_i(b)),
\end{equation}
\begin{equation}
\label{SmxRussekMeli}
S_{m}(x)= 2^{ \{(m-1)/2 \} } \pi^{\{m/2 \} } x^{(3m-2)/2} / (3m-2)!!
\end{equation}
\begin{equation}
\label{etaRussekMeli}
\eta_k(b)= \frac{T(b)}{I_1} - \sum^{k}_{i=1}\frac{I_{k}}{I_1},
\end{equation}
Here $\binom{N'}{m}$ denotes the binomial coefficient,
$\{a\}$ is the integer part of $a$. Double factorial means
$m!!=m(m-2)(m-4)\ldots p$ with $p=1$ for odd $m$
and $p=2$ for even $m$ correspondingly,
and $S_m(x<0)=0$ by definition.

The probabilities $P_m(b)$ satisfy the normalization condition at
every impact parameter $b$
\begin{equation}
\label{PmSum1}
\sum_{m=1}^{N'}P_{m}(b)=1,
\end{equation}
as it follows from the equations
(\ref{PmbRussekMeli})-(\ref{etaRussekMeli}). The variable $N'$
here means the number of active electrons considered to be
ionized. Actually, $N'$ in equations (\ref{PmbRussekMeli}),
(\ref{PmSum1}) should be equal to the number of projectile
electrons $N$, but due to the rapid convergence of the sum
(\ref{PmSum1}) it can be cut off at $N'\le N$ value.

Using the deposited energies $T(b)$ calculated in the DEPOSIT code
and the ionization potentials~$I_{k}$ of the projectile ion, the
$m$-fold EL cross sections $\sigma_m(v)$ can be calculated
straightforwardly by applying equation (\ref{sigmam}).

\section{Numerical procedure}
\label{NumericalProc2}

From the numerical point of view the challenge 
for the evaluation of the deposited energy~(\ref{Tb3DIntegral})
is a three-dimensional integral over 
the projectile-ion coordinate space.
For a given $\gamma$-shell it can be written
in general form as
\begin{equation}
\label{EbNumIntgrl}
T_{\gamma}(b)=\int_{0}^{\infty}dr \int_{-1}^{1}dx
\int_{0}^{2\pi}d\varphi
R(r) F(b,r,x,\varphi).
\end{equation}
For simplicity, $\gamma$-notation is skipped in the
right-hand side of the integral.
Parameter $b$ is fixed, and functions
$R(r)$ and $F(b,r,x,\varphi)$
are smooth, real and do not have any singularities.
Function $F(b,r,x,\varphi)$ is nothing but $\Delta E_{\gamma}(p)$
given by the expressions (\ref{EfinalLow}) and~(\ref{EfinalHigh}),
and the radial function $R(r)$ has a Slater-type behaviour
and represents the electron-density contribution~(\ref{rhodencity})
\begin{equation}
\label{RadialSlaterFunc}
R(r)=N_{1} C^{2}_1 r^{2 \mu} e^{-2 \beta r}.
\end{equation}
Here $C_1$ is a normalization factor
from the equation (\ref{rhodencity}),
$N_1$ is the number of electrons in given $\gamma$-shell.

In the expression (\ref{EbNumIntgrl}) all three integrals
over the variables $r,x$ and $\varphi$
have a constant upper and lower limits.
It means that each can be split out and
considered separately like one-dimensional
integral with a given set of fixed parameters, namely
\begin{equation}
\label{AnglePhiIntegral}
F_{brx}(b,r,x)=\int_{0}^{2\pi} d\varphi F(b,r,x,\varphi),
\end{equation}
\begin{equation}
\label{CosX11Integral}
F_{br}(b,r)=\int_{-1}^{1}dx F_{brx}(b,r,x),
\end{equation}
\begin{equation}
\label{RadialIntegral}
T_{\gamma}(b)=\int_{0}^{\infty}dr R(r) F_{br}(b,r).
\end{equation}

\subsection{Integration over the angles}

In the integral (\ref{AnglePhiIntegral})
the dependency on $\varphi$ is expressed
only by $\cos^2\!\varphi$ as it follows from
the equation (\ref{pbtetphiBound}).
Due to the symmetry of
the $\cos^2\!\varphi$ function, implementation of the
Simpson's rule~\cite{NumMethIngSci}, \cite{KrylovInt}
to the integral (\ref{AnglePhiIntegral})
can be done in the following way
\begin{equation}
\label{KrylovAngle2Int}
\int_{0}^{2\pi}f(\cos^2\!\varphi) d \varphi\approx
\frac{2\pi}{3 N_{c}}
\left(
S_{\varphi}+f_{0}+4f_{1}+f_{N_{c}}
\right),
\end{equation}
\begin{equation}
\label{NcEq31}
S_{\varphi}=\sum_{k=1}^{N_{c}/2}
\left(
4f_{2k+1}+2f_{2k}
\right),
\qquad
f_{k}=f(\cos^2 \frac{\pi k}{2 N_c}),
\end{equation}
where $N_c$ must be even and $f(\cos^2 \varphi)=F(b,r,x,\varphi)$.
Discrepancy estimation for eq.~(\ref{KrylovAngle2Int})
is given in ref.~\cite{KrylovInt}.

Integral~(\ref{CosX11Integral}) can be done
by the following rule 
\begin{equation}
\label{Krylov2IntLab}
\int^{1}_{-1}f(x)dx \approx
2\frac{f(-1)+f(1)}{(n+1)(n+2)}
+\sum_{k=1}^{n} A_k f(x_k),
\end{equation}
\begin{equation}
\label{Krylov2AkoeffLab}
A_k=\frac{8(n+1)}{(n+2)(1-x_{k}^{2})^2
[(P^{(1,1)}_{n}(x_k))^{'}]^2},
\end{equation}
with an error estimation
given in ref.~\cite{KrylovInt}.
Here, $P^{(1,1)}_n(x)$ is a Yacobi polynomial,
$x_k$ is the $k$-th root of the polynomial $P^{(1,1)}_n(x)$,
$f(x)=F_{brx}(b,r,x)$.
For the choice of the $n=53$
numeric error is equal to 
$\lambda \cdot 10^{-15}$, where
numeric factor $\lambda$ depends on a given system,
i.e. the function $F_{brx}(b,r,x)$.

To speed up the computing procedure
based on formulas (\ref{Krylov2IntLab}), (\ref{Krylov2AkoeffLab})
preliminary calculations of the Yacobi polynomial roots $x_k$
and coefficients $A_k$ have been done.
These two data sets have been directly built in the present code.

\subsection{Integration over the radius}

Due to the Slater-type behaviour of the
radial function (\ref{RadialSlaterFunc}),
integrand $R(r) F_{br}(b,r)$ decreases exponentially.
The main contribution to the integral (\ref{RadialIntegral})
function $R(r) F_{br}(b,r)$ makes in the range $[0,r_{\max}]$.
The infinity limit can be cut off in the $r_{\max}$ point.
The value of $r_{\max}$ is an input parameter,
which is ordinary taken approximately
equal to~$100$, depending on the projectile ion size.
Input parameters are described in Section~\ref{ExamplesIO3}.

To do integral~(\ref{RadialIntegral}) the logarithmic
grid with an input parameter $N_{grid}$ is used
\begin{equation}
\label{RgridDef}
r_{i}=r_{\max}  e^{(i - N_{grid})/t_0},
\qquad
0 \le i \le N_{grid}, \qquad t_0=32,
\end{equation}
and a Newton-Cotes quadrature formula
for the seven-points interval
\begin{equation}
\label{NewtonKotessAppr}
\begin{split}
I^{i+1}_{i}=\frac{x_{i+1}-x_{i}}{6 \cdot 140}
[
41 f(x_{i})+ 216 f(x_{i+\frac{1}{6}})+ 27 f(x_{i+\frac{2}{6}})+ \\
272 f(x_{i+\frac{3}{6}})+ 27 f(x_{i+\frac{4}{6}})+
216 f(x_{i+\frac{5}{6}})+ 41 f(x_{i+1})
],
\end{split}
\end{equation}
is applied for every range $[r_{i},r_{i+1}]$
(see ref.~\cite{Abramowitz},  p.886, eq.~25.4.16).
Here $x_i=r_i$ and points in the interval $[x_{i},x_{i+1}]$ are distributed uniformly.

In this way integral (\ref{RadialIntegral}) can be done by the following rule
\begin{equation}
\label{TbImplIiip1}
\int_{0}^{r_{\max}} f(r)dr\approx
\sum_{i=0}^{N_{grid}-1} I^{i+1}_{i}.
\end{equation}

\subsection{The total cross-section calculations}
\label{SigTotSect}

\begin{figure*}
\includegraphics[width=16cm]{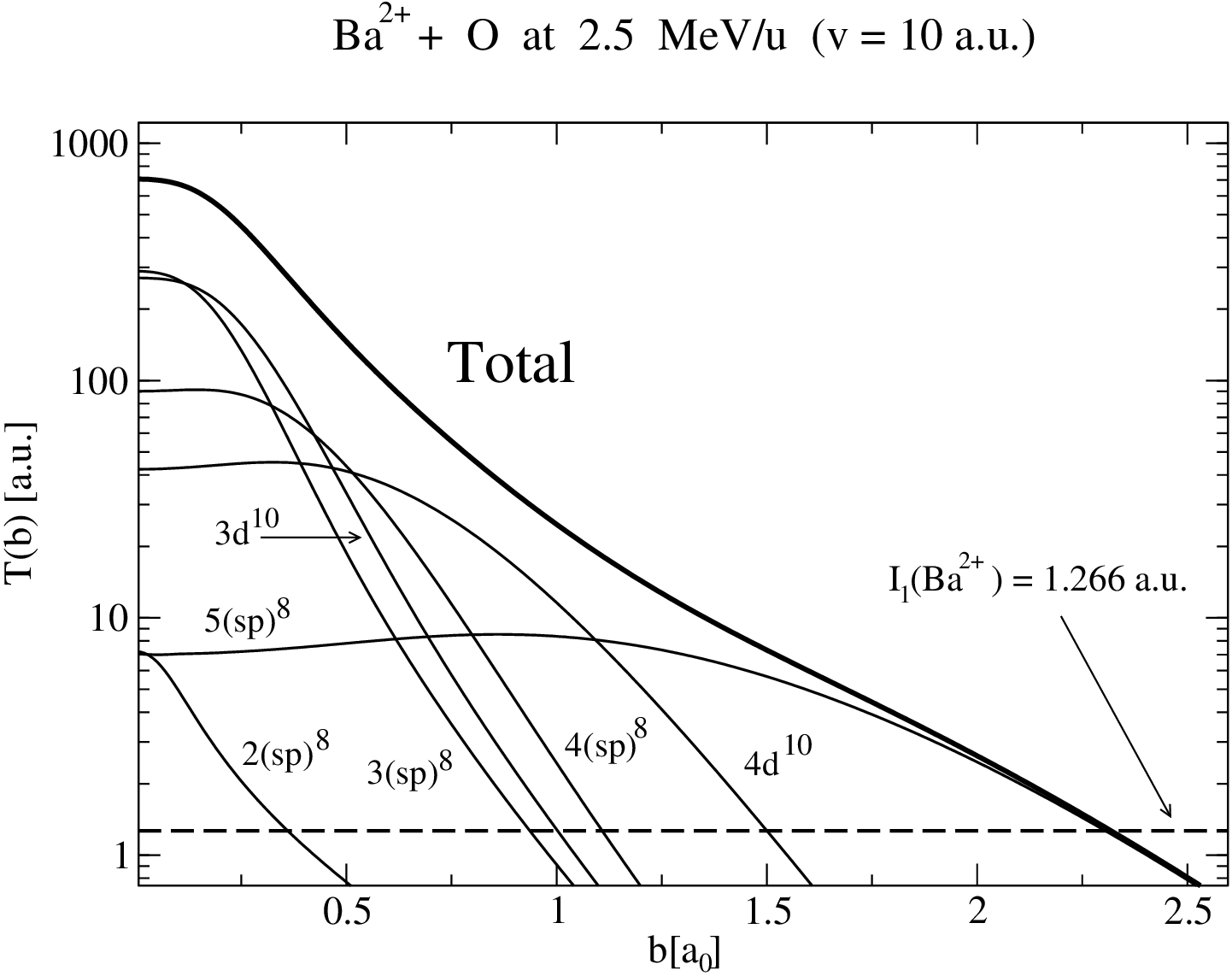}
\caption{
\label{fig1Lbl}
Energy deposition $T(b)$ (in a.u.) to $Ba^{2+}$ ion
by an $O$ atom at a collision energy $E=2.5$ MeV/u (v = 10 a.u.)
as a function of the impact parameter $b$,
calculated with the DEPOSIT code.
Energies deposited onto different electronic
(Slater) shells of $Ba^{2+}$ are shown
together with the total energy deposition.
The horizontal dashed line $I_1=1.266$ a.u. (34.45 eV)
corresponds to the
first ionization potential of $Ba^{2+}$
and shows the minimal energy deposition
required for the ionization process.
The notation for the Slater shells, for example,
$2(sp)^8$ means the electron configuration of the $2s^22p^6$ shell.
}
\end{figure*}

The total-cross section $\sigma_{tot}$ is calculated
from equation~(\ref{sigmatotEq}), where
the $b_{\max}$ parameter is taken as a numerical solution
of equation (\ref{bmaxEquation}).
Function $T(b)$ monotonically decreases 
with increasing of parameter $b$ 
near the $b_{\max}$ point
(a typical example of $T(b)$ behaviour is given
in Figure~\ref{fig1Lbl}).

To solve equation~(\ref{bmaxEquation})
under these conditions the bisection method
is applied within a given range $[b_1,b_2]$,
where $b_1$ and $b_2$ are input parameters.
The bisection procedure is stopped when the condition
\begin{equation}
\label{BisectCond}
b^{(i)}_2 - b^{(i)}_1 < \varepsilon, \qquad \varepsilon = 10^{-3},
\end{equation}
is satisfied, where $i$ is the number of iteration.
Then the quadratic interpolation procedure is used to
find the local minimum of the $(T(b)-I_1)^2$ function
near the $b_{\max}$ value using three abscissa points
\begin{equation}
\label{Parabolamin}
b_0-\varepsilon/4, \quad b_0, \quad b_0+\varepsilon/4,
\end{equation}
where $b_0=(b^{(i_0)}_1 + b^{(i_0)}_2)/2$, $i_0$ is the final
bisection step controlled by the condition~(\ref{BisectCond}).
Once the $b_{\max}$ parameter is determined,
the $\sigma_{tot}$ value is calculated by
applying the equation~(\ref{sigmatotEq}).

\subsection{The m-fold cross-section calculations}
\label{SigmFoldSect}

The $m$-fold cross sections $\sigma_{m}$ are
calculated by formula~(\ref{sigmam}),
where the integration is done
from $0$ up to $b_{\max}$. To do integral~(\ref{sigmam})
one, first of all, needs to calculate $T(b)$ values
in every point of the integration grid
as it follows from equations~(\ref{PmbRussekMeli})
and (\ref{etaRussekMeli}).

The grid is taken with a fixed step $h\le0.01$ on the
$[0,b_{\max}]$ range. The number of points
\begin{equation}
\label{ProbabNpgrid}
N_{p} = [(1+\{b_{\max}/0.01\})\,\,
\mbox{\textbf{div}} \,\, 2] \cdot 2,
\end{equation}
and the grid points are
\begin{equation}
\label{ProbabGrid}
x_j=jh, \quad
j=0,1 \ldots N_p, \quad
h=b_{\max}/N_{p}.
\end{equation}
The integer part of $a$ in equation~(\ref{ProbabNpgrid}) is
labeled as $\{a\}$, and the integer division operation is noted as \textbf{div}.

The Simpson rule
\begin{equation}
\int_{0}^{b_{\max}} f(x) dx \approx
\frac{h}{3}\left[
f(x_0) +F_2 + F_4 +f(x_{N_p})
\right],
\end{equation}
\begin{equation}
F_2 = 2\sum_{j=1}^{N_p/2-1} f(x_{2j}),
\qquad
F_4 = 4\sum_{j=1}^{N_p/2} f(x_{2j-1})
\end{equation}
is applied to do integration~(\ref{sigmam}) on
grid~(\ref{ProbabGrid}).

The probabilities $P_m(b)$ are implemented in correspondence
to equations~(\ref{PmbRussekMeli}) -- (\ref{etaRussekMeli})
with the following boundary conditions
\begin{equation}
P_m(b < \epsilon_0)=
\begin{cases}
1, \quad  m=1, \\
0, \quad 1< m \le N',
\end{cases}
\end{equation}
\begin{equation}
P_m(b > M_{\infty})=
\begin{cases}
0, \quad  1 \le m < N', \\
1, \quad m = N',
\end{cases}
\end{equation}
where $\epsilon_0=10^{-12}$ and $M_{\infty}=10^{20}$,
$N'$ is a cutoff value for the sum~(\ref{PmSum1}).

To avoid divergence in the numeric calculations of factorials and
powers of $2$, $\pi$ and $x$ in equations~(\ref{PmbRussekMeli}) and~(\ref{SmxRussekMeli}) the following
approach to $P_m(b)$ calculation have been implemented
\begin{equation}
\label{PmExpTransf}
P_m(b)=\left[1+ \sum_{i=1, i\ne m}^{N'}\exp(\Lambda^m_{i}(b)) \right]^{-1},
\end{equation}
\begin{equation}
\Lambda^m_{i}(b)=
S^{m,i}_{1}+S^{m,i}_{2}+
S^{m,i}_{3}+S^{m,i}_{4}+
S^{m,i}_{5}(b)
\end{equation}
\begin{equation}
S^{m<i}_1=\sum_{k_1 = N' - i + 1}^{N' - m} \ln k_1-
\sum_{k_2=m + 1}^{i} \ln k_2,
\end{equation}
\begin{equation}
S^{m>i}_1=\sum_{k_1=i + 1}^{m} \ln k_1-
\sum_{k_2 = N' - m + 1}^{N' - i} \ln k_2,
\end{equation}
\begin{equation}
\label{FactSumOddEven}
S^{m,i}_2=
\sum_{k_1+=2}^{3m-2} \ln k_1-
\sum_{k_2+= 2}^{3i-2} \ln k_2,
\end{equation}
\begin{equation}
S^{m,i}_{3}=\left( \left \{ (i - 1)/2  \right \} -
\left \{ (m-1)/2 \right \} \right ) \ln 2,
\end{equation}
\begin{equation}
S^{m,i}_{4}=\left( \left \{i/2\right \} -
\left \{ m/2 \right \}
\right )
\ln\pi,
\end{equation}
\begin{equation}
S^{m,i}_{5}(b)=(3i/2-1) \ln \eta_{i}(b)- (3m/2-1) \ln \eta_{m}(b).
\end{equation}

Summation in the first sum of equation~(\ref{FactSumOddEven})
over $k_1$ is done for all even positive $k_1 \le 3m-2$, if $m$ is even,
and for all odd positive $k_1$, if $m$ is odd, as well as summation
over even/odd positive $k_2$ for even/odd $i$ in the second
sum of equation~(\ref{FactSumOddEven}) correspondingly.

\begin{figure*}
\includegraphics[width=16cm]{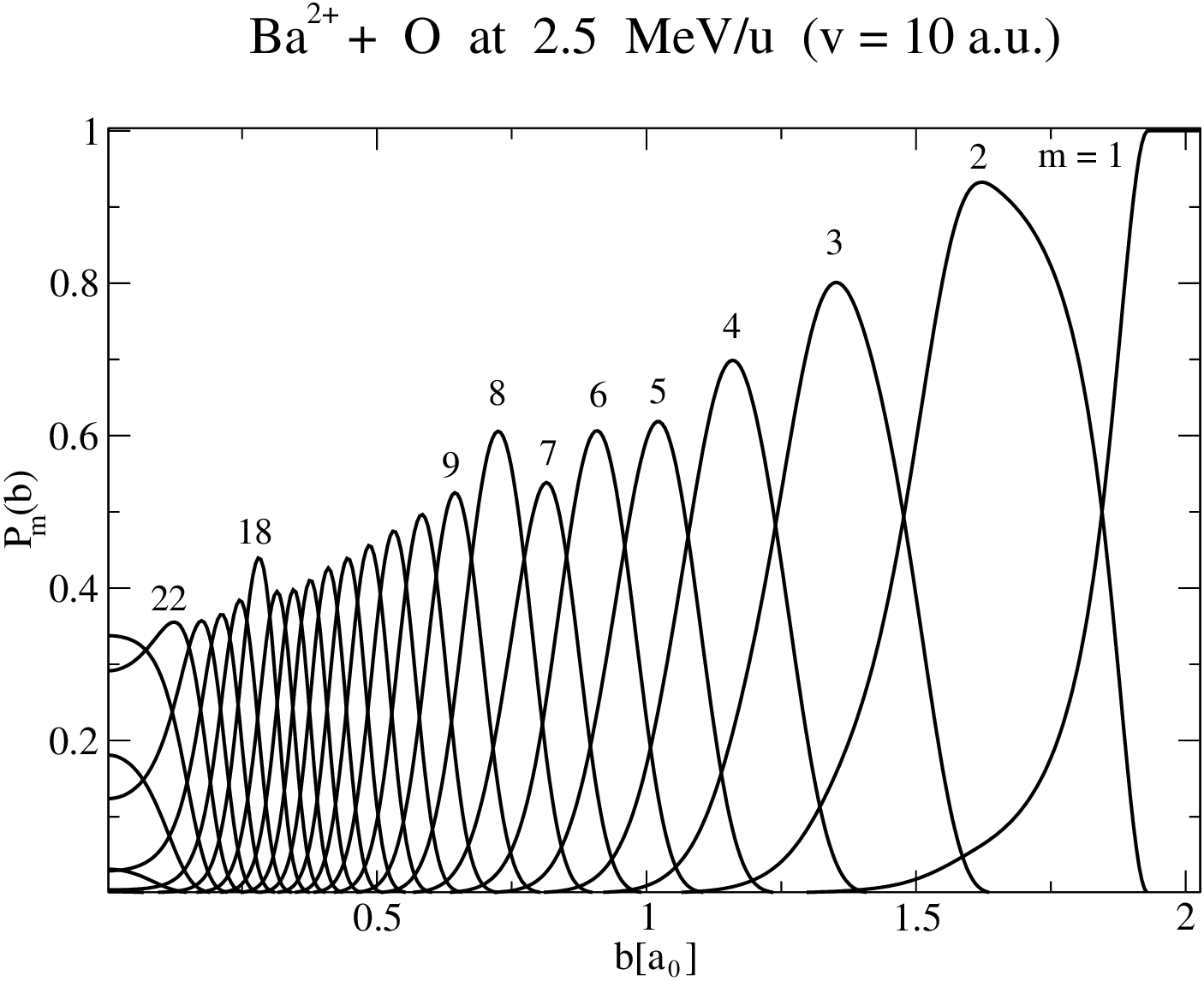}
\caption{ \label{fig2Lbl} Calculated probabilities $P_m(b)$ for
the $m$-fold ionization of $Ba^{2+}$ ions colliding with $O$ atoms
at energy $E=2.5$ MeV/u (v = 10 a.u.) as a function of the impact
parameter $b$, eqs.~(\ref{PmbRussekMeli})-(\ref{etaRussekMeli}). }
\end{figure*}

Integral (\ref{sigmam}) for every $\sigma_m$ is done
on the same grid~(\ref{ProbabGrid}) $[0,b_{\max}]$.
An example of the $P_m(b)$ probabilities
is given in Figure~\ref{fig2Lbl}.

\subsection{Parallelization}

Integral (\ref{Tb3DIntegral}) representing
the deposited energy $T(b)$ after all implementation
steps is finally reduced to the summation
rule~(\ref{TbImplIiip1}).
It can be broken down into the
$N_{core}$ subsumes where $N_{core}$ is the number
of MPI (Message Passing Interface) threads (cores).
For parallel version of the present
code the OpenMPI library is used~\cite{openmpiref}.

In the parallel version integration in~(\ref{TbImplIiip1}) for
a given $q$-th core, $0\le q < N_{core}$, is done by the rule
\begin{equation}
\label{InPartSum}
I(q)=\sum_{i=i_1[q]}^{i_2[q]-1} I^{i+1}_{i},
\end{equation}
where
\begin{equation}
i_1[q]=q\left(\left\{\frac{N_{grid}}{N_{core}}\right\}+1\right),
\end{equation}
\begin{equation}
i_2[q]=
\begin{cases}
i_1[q+1], \quad i_1[q+1] \le N_{grid}, \\
N_{grid}, \quad otherwise.
\end{cases}
\end{equation}
The integer part of $a$ is labeled as $\{a\}$.
After the partial summations~(\ref{InPartSum}) on every core
the final result
\begin{equation}
\sum_{q=0}^{N_{core}-1}I(q)=
\sum_{i=0}^{N_{grid}-1} I^{i+1}_{i}
\end{equation}
is collected by calling the
\texttt{Allreduce} MPI function.

As one can see the parallelization
having done in the way~(\ref{InPartSum})
does not harm the rule~(\ref{NewtonKotessAppr}).
This remarkable property
combines high accuracy with high performance.

\section{Input/output}
\label{ExamplesIO3}

\subsection{Input}
\label{SubSecInp}
All input parameters for the program  are set up by means of
one input text file with the proper keywords.
Every keyword can be put in arbitrary place
of the input file without
a definite compliance of the order.
The program is case-sensitive to the keywords.
Comments can be added in
$C$/$C$++ programming language style.

The keywords are: \texttt{Vi} - velocity
of the projectile in frame of the
neutral atomic target given in atomic units,
\texttt{ZA} - atomic charge and
\texttt{RA} - effective radius of the atomic target,
see equations~(\ref{EfinalLow}), (\ref{EfinalHigh}) and~(\ref{NeffCase}).
Each of these parameters should be followed by
one or several spaces and a real number (the value of the parameter).
For example, for the $Ba^{2+}+O$ collision
at the energy $E=2.5$~MeV/u they are
\begin{verbatim}
 Vi     10.0
 Za      8.0
 Ra      1.239652
\end{verbatim}
Two values $A_1$ and $A_2$ for the equation~(\ref{DiracUr})
are given after the keyword \texttt{A\_exp}.
Three values $\alpha_1$, $\alpha_2$, $\alpha_3$
for the equation (\ref{DiracUr}) are given after
the keyword \texttt{alf\_exp}.
\begin{verbatim}
A_exp     0.0625   0.9375
alf_exp   14.823   2.0403   0.0
\end{verbatim}

First integer value after the keyword \texttt{shells}
sets up the number of the ion $\gamma$-shells.
The number of electrons
$N_{\gamma}$, the values of the coefficients
$C_{1\gamma}$, $\mu_{\gamma}$, $\beta_{\gamma}$
for the density (\ref{RadialSlaterFunc}) and binding energy $I_{\gamma}$
must be put in as well.
For every $\gamma$-shell it is more convenient
to set up the five parameters from a new line.
Rules for calculation of the $C_{1\gamma}$,
$\mu_{\gamma}$, $\beta_{\gamma}$
parameters are given in
the references~\cite{nimb09},~\cite{ShevSlaterPrm2001}.
The binding energies $I_{\gamma}$ are put in
electron-volts~(eV). The values of $I_{\gamma}$ can be
found in \cite{Desclaux},~\cite{IbRef2}-\cite{LoFite}.
\begin{verbatim}
 Shells 7
//N      C1      mu    beta     I(eV)
  8     7.778   4.0   2.5625    33.11   // 5sp8
 10   121.07    3.5   4.8143   106.044  // 4d10
  8   956.5335  3.5   8.0714   215.52   // 4sp8
 10  2252.830   3.0  11.6167   801.35   // 3d10
  8  5404.971   3.0  14.9167  1152.6    // 3sp8
  8  3951.539   2.0  25.925   5542.13   // 2sp8
  2   831.405   1.0  55.7    37455.41   // 1s2
\end{verbatim}

The keyword \texttt{b\_range} gives $b_{\min}$, 
$b_{\max}$, and $\Delta b$ for calculating and plotting $T(b)$.
\begin{verbatim}
// T(b) curve
   b_range  0.0  3.0  0.01
\end{verbatim}
The step parameter $\Delta b$ does not need to be the range
devided by an integer; $b$-points are chosen by the rule
$b_i=b_{min} + i \Delta b$, $i=0,1,2\ldots k$,
$b_{k} \le b_{max}$, $b_{k+1} > b_{max}$.

To calculate the total cross section $\sigma_{tot}$ one needs
to give a keyword \texttt{Sigma\_tot}
with three parameter values:
bisection start range $[b_1, b_2]$
(see Section~\ref{SigTotSect})
and the first ionization potential~$I_1$~[eV].
\begin{verbatim}
// Total cross section
   Sigma_tot  0.0  11.0    34.45
\end{verbatim}

To calculate $m$-fold cross sections
$\sigma_m$, where $1\le m \le N'$, one needs
to give a keyword \texttt{Sigma\_m\_fold} with
$N'+1$ values:
the value of $N'$ parameter and
$N'$ values of $m$-th ionization
potentials $I_m$~[eV]
(see equations~(\ref{sigmam})-(\ref{PmSum1}) and
Section~\ref{SigmFoldSect}).
\begin{verbatim}
// m-fold cross section
   Sigma_m_fold  30
   34.45   48.40   62.35   76.30   92.53
   107.1   139.2   155.7   232.2   266.0
   299.8   333.6   367.4   401.2   437.8
   472.0   506.2   540.5   672.6   710.3
   747.9   785.6   839.2   878.2   975.9
  1016.0  1586.0  1688.0  1790.0  1892.0
\end{verbatim}
One important point is that in order to calculate $m$-fold
cross sections one must also calculate
the $b_{\max}$ value from equation~(\ref{bmaxEquation}).
It is needed to cut off the integration
grid in equation~(\ref{sigmam}). Thus,
the \texttt{Sigma\_m\_fold} keyword must be given
in the text file with the keyword
\texttt{Sigma\_tot} and all their data.

The remaining parameters should be kept with the default values.
\begin{verbatim}
 rgrid     70.0  600  30
 ksmear    3.0
 cosN      54
\end{verbatim}
Parameter \texttt{rgrid} sets up the
$r_{\max}$, $N_{grid}$ and $t_0$ values for equation~(\ref{RgridDef}).
The $r_{\max}$ corresponds to the physical infinity
for the atomic radial variable.
Parameter \texttt{ksmear} fixes the $k$ value in equation~(\ref{kSmear}) and
parameter \texttt{cosN} sets up the number of points $N_c$
in equations~(\ref{KrylovAngle2Int})-(\ref{NcEq31}).
Value of the \texttt{cosN} parameter must be only even.

\subsection{Output}
\label{SubSecOut}
The name of the input file is passed to the program input
with the name of the output file
via command line when the program starts.
The output of the program is sent to console.
\begin{verbatim}
> deposit file_name1_inp file_name2_out
\end{verbatim}

The program writes a text file
with the deposited energy called
\texttt{energy\_Tb\_[file\_name2\_out]},
if the \texttt{b\_range} keyword is given.
The file consists of several columns.
The first column contains the $b$-values,
the second column is the $T(b)$ energy
and the rest of the columns are the deposited energies
for every $\gamma$-shell.

The program also writes a text file
with the probabilities $P_m(b)$
called \texttt{probability\_Pm\_[file\_name2\_out]},
if the \texttt{Sigma\_m\_fold} keyword is given.
The file consists of several columns.
The first column is the $b$-values in eq.~(\ref{PmbRussekMeli})
and the other ones are values of the $P_m(b)$
probabilities for the corresponding $b$ points
and $1 \le m \le N'$.

All input parameters are printed out to the console
(for the user's control) and the program starts the calculations.

\begin{verbatim}
 ****  DEPOSIT CODE (VERSION 1.27/P) ****

 Number of MPI Threads = 8

 Input reading from file:  ba2+_o.txt

 Vi = 10.0  Za = 8  Ra = 1.23965

 A_exp:     0.06250   0.93750   0.00000
 alf_exp:  14.82300   2.04030   0.00000

 Shells = 7
   N        C1     mu   beta      I[eV]   .
  8.0      7.778  4.0   2.562      33.11  .
 10.0    121.070  3.5   4.814     106.04  .
  8.0    956.534  3.5   8.071     215.52  .
 10.0   2252.830  3.0  11.617     801.35  .
  8.0   5404.971  3.0  14.917    1152.60  .
  8.0   3951.539  2.0  25.925    5542.13  .
  2.0    831.405  1.0  55.700   37455.41  .

 .    I[au]      u      Neff
 .     1.22    1.56     ---
 .     3.90    2.79     ---
 .     7.92    3.98     ---
 .    29.45    7.67     ---
 .    42.36    9.20     ---
 .   203.67   20.18   0.0511
 .  1376.46   52.47   0.0019

 (r_max =  70.0   Ngr = 600   t0 = 30.0
  ksmear = 3.0   cosN = 54)

 Slater w.f. normalization test:
  P_1:  1.0000
  P_2:  1.0000
  P_3:  1.0000
  P_4:  1.0000
  P_5:  1.0000
  P_6:  1.0000
  P_7:  1.0000
\end{verbatim}
According to the equations~(\ref{EfinalLow})
and (\ref{EfinalHigh})
the $N_{\mbox{\footnotesize{eff}}}^{(\gamma)}$ parameter
is used only if $v \le u_{\gamma}$.
In this case the $N_{\mbox{\footnotesize{eff}}}^{(\gamma)}$
values are printed out in the \texttt{Neff} column.
Otherwise, when $v > u_{\gamma}$, only a dashed line is printed.

For calculations of the deposited energy $T(b)$
the program prints out the $b$-mesh information
and two columns ($b$-values and $T(b)$-values).
\begin{verbatim}
 b_grid:  [0.000; +0.010; 3.000]  Npoints = 301

 Deposited energy T(b) calculation
      b          T(b)
  0.000000    706.586257
  0.010000    706.333016
  0.020000    705.566015
  . . . . . . . . . . . .
  2.980000      0.216640
  2.990000      0.210448
  3.000000      0.204421
 Output T(b) into the file:
 energy_Tb_ba2+_o_out.txt
\end{verbatim}

For the $\sigma_{tot}$ calculations
the program prints out the input parameters,
bisection search history~(\ref{BisectCond}),
and interpolation data~(\ref{Parabolamin}).
Finally, it prints out the $b_{\max}$
and $\sigma_{tot}$ values.
\begin{verbatim}
 Total cross-section calculation
 I1 =     1.2660 a.u.
 b_min =  0.0000   b_max = 11.0000

 Bisection search:
      b       T(b) - I1
  0.000000    705.320243
 11.000000     -1.266014
  5.500000     -1.265965
  2.750000     -0.851313
  1.375000      8.266428
  2.062500      1.021403
  2.406250     -0.245276
  2.234375      0.280613
  2.320312     -0.005746
  2.277344      0.131173
  2.298828      0.061201
  2.309570      0.027356
  2.314941      0.010713
  2.317627      0.002461
  2.318970     -0.001648
  2.318298      0.000405

 Interpolate:
  2.318384      0.000143
  2.318634     -0.000622
  2.318884     -0.001386

 b_total     =    2.318431
 Sigma_total =   16.886438 a.u.
 Sigma_total =    4.728683e-16 cm2
\end{verbatim}

For the $m$-fold cross-section $\sigma_{m}$ calculations,
it prints out input values of ionization
potentials, $b$-grid data for the integral~(\ref{sigmam})
and values of the deposited energy.
Once $T(b)$ energy is calculated, the
$\sigma_{m}$ data are evaluated and printed out.
To check out the calculation
the sum rule (\ref{sigmatotEq}) is applied to the
$\sigma_{m}$ values and printed after
the partial cross sections.
\begin{verbatim}
 m-fold cross-sections calculation
  I_1 =     1.2660 a.u. =       34.450 eV
  I_2 =     1.7787 a.u. =       48.400 eV
  . . . . . . . . . . . . . . . . . . . .
 I_28 =    62.0329 a.u. =     1688.000 eV
 I_29 =    65.7813 a.u. =     1790.000 eV
 I_30 =    69.5297 a.u. =     1892.000 eV

 m-fold Probability grid:
 N_points  = 232
 b_max     =   2.3184
 grid_step =   0.0100
 Save points:
     b          T(b)
  0.0000      706.5863
  0.0100      706.3334
  0.0200      705.5674
  0.0300      704.2550
  . . . . . . . . . .
  2.2885        1.3606
  2.2984        1.3284
  2.3084        1.2969
  2.3184        1.2660

  m-fold Cross-sections:
    m       a.u.        cm2
    1    6.512001   1.823545e-16
    2    3.688431   1.032866e-16
    3    1.949328   5.458672e-17
    4    1.146097   3.209396e-17
    5    0.750239   2.100883e-17
   . . . . . . . . . . . . . . .
   26    0.000030   8.382028e-22
   27    0.000000   2.943365e-25
   28    0.000000   1.303180e-30
   29    0.000000   2.553841e-40
   30    0.000000   0.000000e+00
  sum   16.886438   4.728683e-16

 Output Pm(b) into the file:
 probability_Pm_ba2+_o_out.txt

 Start time: Sun May 20 14:18:36 2012
 Stop  time: Sun May 20 14:27:55 2012
\end{verbatim}

\begin{figure*}
\includegraphics[width=16cm]{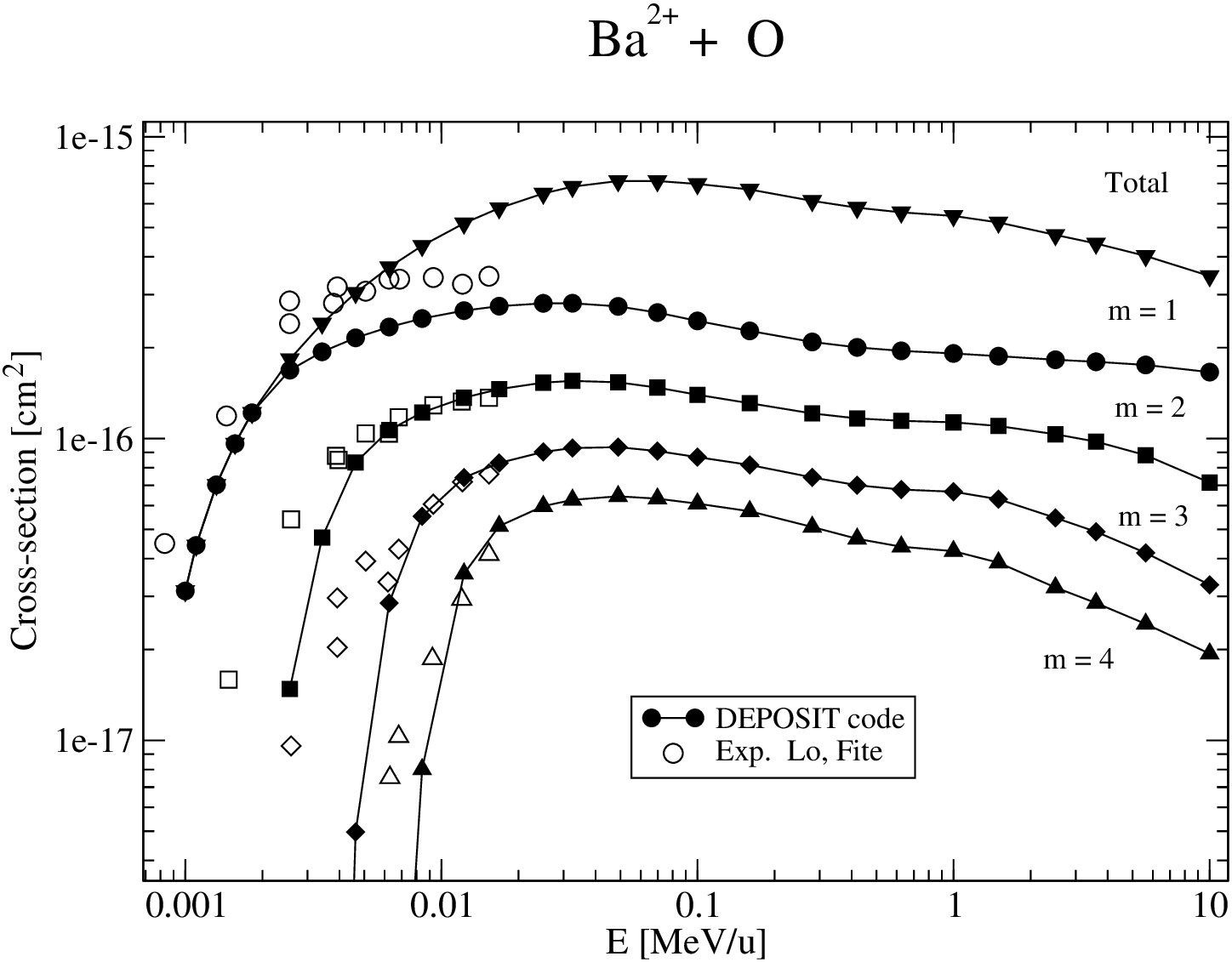}
\caption{ \label{fig3Lbl} The total and $m$-fold electron-loss
cross sections of $Ba^{2+}$ ions colliding with $O$ atoms as a
function of ion energy for ejection of $m=1$, $2$, $3$, $4$
electrons calculated by the DEPOSIT code (solid symbols).
Experimental data from \cite{LoFite} are available only for the
$m$-fold cross sections (open symbols).
At energies $E<2$~keV/u only one-electron losses occur,
that is why total and  one-electron loss cross sections are the same in this region.}

\end{figure*}

Example of the calculated $m$-fold cross sections
versus collision energies
is given in Figure~\ref{fig3Lbl}.

\section{Conclusions}
\label{ResultsDiscSect} A description of the DEPOSIT computer code,
based on the energy-deposition model, is presented. This approach gives
an agreement with experimental data within a factor of two.
Numerical implementation of the model is given as well. The
algorithms realized in the program allows one to simulate
ion-atomic collisions at low and intermediate energies for
arbitrary ion-atomic colliding systems. The code is parallelized and
can be run in single and parallel modes (on a user PC or supercomputer cluster). 
The computational speed is proportional to the number of cores.

\section*{Acknowledgements}
\label{AcknowledgeSect}

The author is very grateful to V.~P.~Shevelko for valuable
discussions and advice during the preparation of this work. The
author is also grateful to O.V.~Ivanov and P.~Th\"onstrom
for helpful discussions of numeric questions.

The computations were performed on resources  provided
by the Swedish National Infrastructure for Computing (SNIC)
at Uppsala Multidisciplinary Center
for Advanced Computational Science (UPPMAX)
under the Project \textit{s00311-8}
and on resources provided by SNIC
at the National Supercomputer Center (NSC)
under the Project \textit{matter2}.


\end{document}